\documentclass[aps,showpacs,superscriptaddress,preprintnumbers,amsmath,amssymb,prb,preprint]{revtex4}
\usepackage{graphicx}
\usepackage{color}
\usepackage{dcolumn}
\usepackage{bm}
\begin{document}
\title{The effect of disorder on quantum phase transition in the double layered ruthenates (Sr$_{1-x}$Ca$_{x}$)$_{3}$Ru$_{2}$O$_{7}$}
\author{Zhe Qu}\email{zhequ@hmfl.ac.cn}\affiliation{Department of Physics, Tulane University, New Orleans, Louisiana 70118, USA}\affiliation{High Magnetic Field Laboratory, Chinese Academy of Sciences, Hefei, Anhui, 230031, China}
\author{Jin Peng}\affiliation{Department of Physics, Tulane University, New Orleans, Louisiana 70118, USA}
\author{Tijiang Liu}\affiliation{Department of Physics, Tulane University, New Orleans, Louisiana 70118, USA}
\author{David Fobes}\affiliation{Department of Physics, Tulane University, New Orleans, Louisiana 70118, USA}
\author{Vlad Dobrosavljevi\`{c}}\affiliation{Department of Physics and National High Magnetic Field Laboratory, Florida State University, Tallahassee, Florida 32306, USA.}
\author{Leonard Spinu}\affiliation{Advanced Material Research Institute and Physics Department, University of New Orleans, Louisiana 70148, USA.}
\author{Zhiqiang Mao}\email{zmao@tulane.edu} \affiliation{Department of Physics, Tulane University, New Orleans, Louisiana 70118, USA}
\date{\today}
\begin{abstract}
(Sr$_{1-x}$Ca$_{x}$)$_3$Ru$_2$O$_7$ is characterized by complex magnetic states, spanning from a long-range antiferromagnetically ordered state over an unusual heavy-mass nearly ferromagnetic (NFM) state to an itinerant metamagnetic (IMM) state. The NFM state, which occurs in the 0.4 $> x >$ 0.08 composition range, freezes into a cluster-spin-glass (CSG) phase at low temperatures [Z. Qu \textit{et al.}, Phys. Rev. B 78, 180407(R) (2008)]. In this article,  we present the scaling analyses of magnetization and the specific heat for (Sr$_{1-x}$Ca$_{x}$)$_{3}$Ru$_{2}$O$_{7}$ in the 0.4 $> x >$ 0.08 composition range. We find that in a temperature region immediately above the spin freezing temperature T$_f$, the isothermal magnetization $M(H)$ and the temperature dependence of electronic specific heat $C_{e}(T)$ exhibit anomalous power-law singularities; both quantities are controlled by a single exponent. The temperature dependence of magnetization $M(T)$ also displays a power-law behavior, but its exponent differs remarkably from that derived from $M(H)$ and $C_{e}(T)$. Our analyses further reveal that the magnetization data $M(H,T)$ obey a phenomenological scaling law of $M(H,T) \propto H^{\alpha}f(H/T^{\delta})$ in a temperature region between the spin freezing temperature $T_{f}$ and the scaling temperature $T_{scaling}$. $T_{scaling}$ systematically decreases with the decease of Ca content. This scaling law breaks down near the critical concentration $x =$ 0.1 where a CSG-to-IMM phase transition occurs. We discussed these behaviors in term of the effect of disorder on the quantum phase transition.
\end{abstract}
\pacs{71.27.+a,74.70.Pq,74.62.En,75.30.Kz} \maketitle

How phase transition is affected by disorder is an important problem extensively studied in condensed matter physics. For a classical phase transition, the system behaves as a clean system at large length scales if the clean correlation length exponent $\nu$ satisfies the criterion $d\nu \leq 2$ (known as the Harris criterion), where $d$ is the space dimensionality. \cite{HarrisCriterion} If this criterion is violated, the system will remain inhomogeneous at all length scales, with the relative strength of the inhomogeneities either remaining finite at all length scales or diverging under coarse graining. Specifically, in addition to these behaviors based on the averaged disorder strength, rare strong fluctuations have been demonstrated to play an important role. In a disordered magnetic system, it is always possible to find "rare regions" which are devoid of any impurities and can locally have a magnetic ordered state even if the bulk system is still in a magnetic disordered phase. Because these rare regions have finite sizes, no true static order can develop and the fluctuations of order parameters are slow. Griffiths \cite{Griffiths} showed that strong fluctuations of these rare regions lead to a singular free energy, which is now known as the Griffiths singularity. In generic classical systems this is a weak effect.

In recent years the focus in this area has shifted to studies of the effect of disorder on quantum phase transitions (QPTs), since the physical properties of many systems of current interest, such as unconventional forms of superconductivity and magnetism, and anomalous non-Fermi-liquid (NFL) behavior, are thought to be controlled over wide thermodynamic ranges by QPTs. \cite{SachdevQPT,SondhiQPT,review214SC,cupQC,Sr327Meta,Sr327QC,Sr327Nematic,NiGa2S4,SrCa214EMIT,StewartNFL,QPTNFL,ScienceHFQPT} Near QPTs order-parameter fluctuates both in space and time. Since the quenched disorder is perfectly correlated in imaginary time direction, rare regions become extended objects, resulting in even slower dynamics and enhancing the Griffiths singularity. \cite{disQPT1,disQPT2,disQPT3} In this scenario, thermodynamics quantities are expected to display "anomalous" power-law singularities controlled by a single continuously varying exponent, $e.g.$ the specific heat $C(T)\propto T^\lambda$, the magnetic susceptibility $\chi(T) \propto T^{\lambda-1}$ and the zero-temperature isothermal magnetization $M(H) \propto H^\lambda$  where $\lambda$ is a non-universal Griffiths exponent and ranges from 0 to 1. \cite{QGpowerlaw1,QGpowerlaw2}

In many itinerant systems close to QPTs, various behaviors reminiscent of the quantum Griffiths singularities have been observed, and their underlying physics is still under debate. \cite{disQPT1,disQPT2,disQPT3} Theoretically it has been suggested that itinerant systems should show quantum Griffiths phenomena similar to those in insulating magnets. \cite{QGpowerlaw1,QGpowerlaw2} However, further theoretical studies suggest that perturbations in itinerant systems could significantly modify this scenario. For example, Landau damping was argued to prevent the tunneling of the rare regions and suppress the Griffiths singularities at low temperatures. \cite{Landaudamping1,Landaudamping2} Moreover, it has also been proposed  that for the Ising case the QPT is fundamentally modified by the overdamping \cite{IsingLandau} while in Heisenberg system the standard quantum critical point (QCP) scenario should hold and the power-law quantum Griffiths singularities could be retained \cite{HeisenbergLandau}. The recent work by Dobrosavljevi\`{c} and Miranda considered the correlation among individual rare regions in itinerant systems and pointed out that such correlations destabilize quantum Griffiths phase, leading to a generic formation of a cluster-spin-glass (CSG) phase preceding uniform ordering. \cite{RKKYQG}
In the meantime, considerate efforts have been devoted to experimentally study the effect of disorder on QPTs. Many experimental results associated with the quantum Griffiths phases have been reported in itinerant systems such as magnetic semiconductors Fe$_{1-x}$Co$_x$S$_2$, \cite{QGsemi} heavy fermion system CePd$_{1-x}$Rh$_x$ \cite{QGHF} and metal alloy Ni$_{1-x}$V$_x$ \cite{QGalloy}. However, the effect of disorder on QPTs was rarely investigated in the transition metal oxides (TMOs).

Here we focus on the double layered ruthenates (Sr$_{1-x}$Ca$_{x}$)$_{3}$Ru$_{2}$O$_{7}$. The end member Ca$_{3}$Ru$_{2}$O$_{7}$ ($x =$ 1) is an antiferromagnet with $T_N =$ 56 K, whose ground state consists of ferromagnetic (FM) bilayers stacked antiferromagnetically along $c$-axis. \cite{Ca327AFM,Ca327Q2D,Ca327magstr,Ca327elastic} Sr substitution for Ca continuously suppresses the $T_N$, from 56 K for $x =$ 1 to 30 K for $x =$ 0.4, and eventually results in the collapse of the AFM order for $x <$ 0.4. \cite{SrCa327flux,SrCa327poly,SrCa327HMNF,SrCa327JPSJ} For 0.4 $> x >$ 0.08, an unusual heavy-mass nearly ferromagnetic (NFM) state is observed, which is characterized by an extremely large Wilson ratio $R_W$ ($e.g.$ $R_{w} \sim$ 700 for $x =$ 0.2). \cite{SrCa327HMNF} Despite considerably strong FM correlations, the system does not develop a long-range order at low temperatures, but freezes into a CSG phase due to the close proximity to a 2D FM instability. \cite{SrCa327HMNF,SrCa327JPSJ} When Ca content is decreased below 0.08, the system enters an itinerant metamagnetic (IMM) state. \cite{SrCa327HMNF} Angle-resolved photoemission spectra of (Sr$_{1-x}$Ca$_{x}$)$_{3}$Ru$_{2}$O$_{7}$ measured at 20 K reveal a qualitative change of electronic properties near $x \sim$ 0.3 $-$ 0.4, which is suggested to be associated with a QPT. \cite{SrCa327ARPES} Specific heat and resistivity measurements down to 0.3 K have proved the existence of non-Fermi-liquid (NFL) behavior near $x =$ 0.1, \cite{SrCa327HMNF,SrCa327EPS} corroborating with the existence of a QPT in this system. Since the system is tuned via chemical substitution, strong disorder is introduced into the system, making the double layered ruthenates (Sr$_{1-x}$Ca$_{x}$)$_{3}$Ru$_{2}$O$_{7}$ a good candidate to explore the effect of disorder on QPT in TMOs.

In this article, we present the scaling analyses of magnetization and the specific heat for (Sr$_{1-x}$Ca$_{x}$)$_{3}$Ru$_{2}$O$_{7}$ in the 0.4 $> x >$ 0.08 composition range.
We find that in a temperature region immediately above the spin freezing temperature T$_f$, the isothermal magnetization $M(H)$ and the temperature dependence of electronic specific heat $C_{e}(T)$ exhibit anomalous power-law singularities; both quantities are controlled by a single exponent.
The temperature dependence of magnetization $M(T)$ also displays a power-law behavior but its exponent differs remarkably from that derived from $M(H)$ and $C_{e}(T)$. Moreover, the magnetization data $M(H,T)$ are found to obey a phenomenological scaling law of $M(H,T) \propto H^{\alpha}f(H/T^{\delta})$ in a temperature region immediately above $T_{f}$. The scaling temperature $T_{scaling}$ below which the scaling equation holds systematically decreases with the decease of Ca content. This scaling law breaks down near the critical concentration $x =$ 0.1 where a CSG-to-IMM phase transition occurs. We discussed these behaviors in term of the effect of disorder on QPTs.

High quality single crystalline samples of (Sr$_{1-x}$Ca$_{x}$)$_{3}$Ru$_{2}$O$_{7}$ were grown using a floating-zone technique. Magnetization measurements were performed using a commercial SQUID magnetometer (Quantum Design) with the magnetic field applied in the basal plane. The specific heat was measured using a thermal relaxation method in a Quantum Design PPMS. The experiments were performed on single crystals that had been characterized in our early works. \cite{SrCa327HMNF,SrCa327EPS,SrCa327Str,SrCa327Ca10}

\begin{figure}[t]
\includegraphics[angle=0,scale=1]{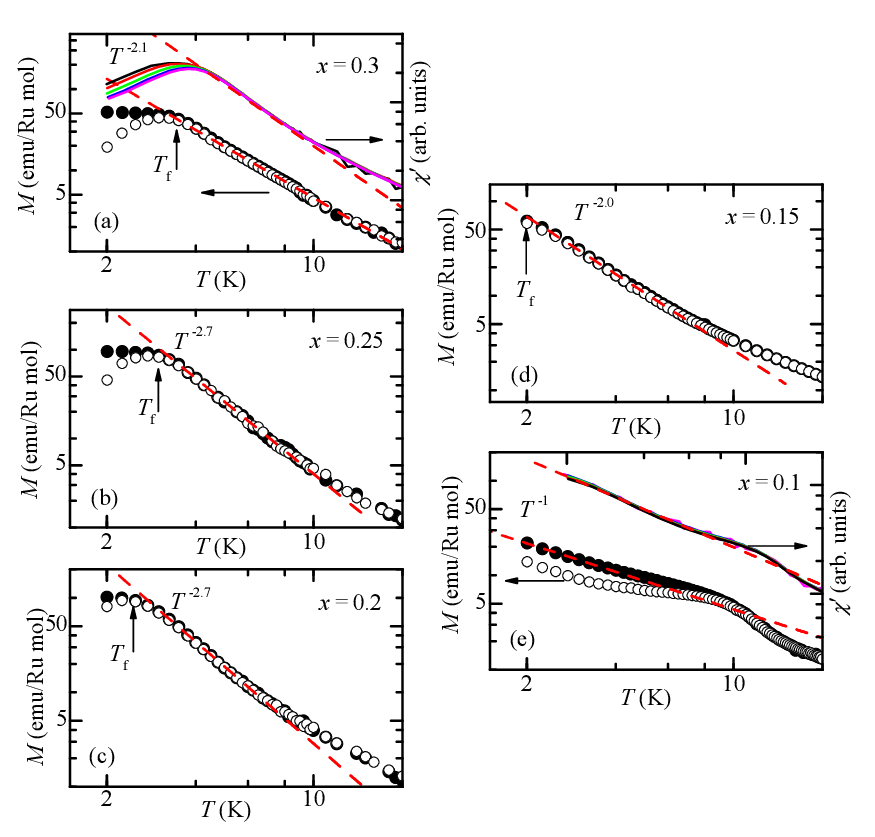}
\caption{(Color online) The temperature dependence of magnetization $M$($T$) under an applied field of 0.005 T with field-cooling (filled symbols) and zero-field-cooling (open symbols) histories for (Sr$_{1-x}$Ca$_{x}$)$_{3}$Ru$_{2}$O$_{7}$ with $x =$ 0.3 (a), 0.25 (b), 0.2 (c), 0.15 (d) and 0.1 (e). The temperature dependence of AC susceptibility measured at 10Hz, 100Hz, 1kHz, and 10kHz for $x =$ 0.3 and 0.1 are also shown in (a) and (e). The magnetic field is applied perpendicular to the $c$-axis. $T_f$ (arrow) marks the freezing temperature of the CSG phase. The dash lines represent fittings to $T^{-\beta}$. The data for $x =$ 0.3 and 0.1 are quoted from from our early works. \cite{SrCa327HMNF,SrCa327Ca10}}\label{fig:MT}
\end{figure}

Figure \ref{fig:MT} shows the temperature dependence of magnetization $M$($T$) taken under an applied field of 0.005 T with field cooling (FC) and zero field cooling (ZFC) histories for typical samples (The data for $x =$ 0.3 and 0.1 are quoted from our earlier works \cite{SrCa327HMNF,SrCa327Ca10}).
The samples freeze into a CSG phase below $T_f$. This is evidenced by i) the irreversibility of $M$($T$) curves between ZFC and FC histories below $T_f$ (shown in Fig. \ref{fig:MT} (a)-(d)), and ii) a frequency dependence of the peak temperature in the real part of AC susceptibility ($e.g.$ see the data for $x =$ 0.3 in Fig. \ref{fig:MT}(a)). For $x =$ 0.1, $T_f$ decreases to $\sim$ 1 K, \cite{SrCa327HMNF} which could not be detected in present measurements. However, distinct irreversibility is observed in $M$($T$) between ZFC and FC histories below $\sim$ 10 K (shown in Fig. \ref{fig:MT} (e)). Given that ferromagnetic correlations are still considerably strong for $x$ = 0.1, as manifested in the large wilson ratio ($\sim$ 110 at 2 K) and that quantum fluctuations are significantly enhanced\cite{SrCa327HMNF}, the irreversible characteristic of $M(T)$ below 10 K may imply formation of ferromagnetic clusters with dynamic nature. Further investigation is needed to demonstrate this point of view.
Additionally, we note that all samples with 0.4 $> x >$ 0.08 show power-law singularities both in DC magnetization and AC susceptibility, $i.e.$ $\varpropto T^{-\beta}$, in a temperature region immediately above $T_{f}$. Those dashed lines in Fig. 1 represent power-law fitting and the power-law exponents derived from fittings are shown in Fig. 4a.

Figure 2 shows isothermal magnetization on log-log scales for $x$ = 0.1 - 0.3. The data have been offset for various temperatures. Apparently, for each sample, $M(H)$ exhibits a liner dependence in a low field region, but a power-law dependence in a high field region, for a given temperature. We have conducted power-law fitting for the data in high-field region, separately, for each temperature. The upper insets in Fig. 2a-2e show the temperature dependence of the power-law exponent $\alpha$. $\alpha$ decreases with decreasing temperature and approaches a constant value ($\sim$ 0.33 - 0.39) at low temperature for all of the samples. This suggests that $M(H)$ follows a similar power-law behavior at low temperatures for different samples. Furthermore, the high field region showing the power-law behavior broadens with decreasing temperature.

We have also defined a crossover field $H^{*}$ to illustrate the transformation from the linear to power-law dependence in $M(H)$. As shown in Fig. 2a-2e, $H^{*}$ is determined as the intersection field between the liner extrapolations of the linear field dependent behavior in the low field region and the power-law field dependence in the high field region (indicated by arrows).
For all samples $H^{\ast}$ decreases with decreasing temperature, following a power-law manner $i.e.$ $H^{\ast} \varpropto T^{\delta}$ (see lower insets to Fig. 2). This means that $M$($H$) is approaching the power-law dependence upon cooling, suggesting that the system develops toward the criticality.
Power-law singularity is also found in the specific heat. The electronic specific heat data were obtained by subtracting the phonon contribution from the total specific heat. \cite{SrCa327HMNF,SrCa327EPS} As shown in Fig. 3, $C_e/T$ shows a power-law temperature dependence for $x$ = 0.15, 0.2, 0.25, and 0.3, $i.e.$ $C_e/T \varpropto T^{\gamma-1}$, in a temperature region right above the freezing temperature of the CSG phase. In contrast, the $x$ = 0.1 sample does not exhibit a similar power-law behavior in $C_e/T$ vs. $T$.\cite{SrCa327HMNF}

Figure 4(a) summarizes the exponents $\alpha$, $\beta$ and $\gamma$ extracted from $M(H)$, $M(T)$ and $C_e(T)$ as described above. It is interesting that the exponent $\gamma$ approximately equals to the exponent $\alpha$ and falls into the range of 0 to 1, for $x =$ 0.15 - 0.3. This power-law singularity appears to be consistent with the prediction of the quantum Griffiths model. \cite{QGpowerlaw1,QGpowerlaw2}
However, several deviations from the prediction of the quantum Griffiths model are noticed. The exponent $\beta$ derived from $M(T)$ is apparently larger than the exponents $\alpha$ and $\gamma$, and does not fall into the expected range of 0 to 1 (see Fig. \ref{fig:exponents} (a)). Meanwhile, the anomalous power-law behaviors do not persist down to the lowest temperature, but are intervened by the formation of the CSG phase. Both deviations suggest that the standard quantum Griffiths model may not be adequate to describe our system.

Such inconsistency might be due to the simplicity of the quantum Griffiths model. As discussed before, only dynamics in a $single$ droplet (rare region) is considered in the standard quantum Griffiths model. However, in a real itinerant system, magnetic moments among droplets are known to interact with each other through the RKKY interaction. The recent theory by Dobrosavljevi\`{c} and Miranda shows that this effect can lead the droplets to freeze in itinerant systems close to QPTs, and gives rise to the generic formation of a CSG phase preceding any uniform ordering. \cite{RKKYQG} Moreover, they pointed out that while a CSG phase generically emerges at the lowest temperature, the anomalous power-law behavior related to the quantum Griffiths phase should be observable in a temperature window above the freezing temperature of the CSG phase. \cite{RKKYQG} This theory can explain the observation of the anomalous power-law singularities at temperatures right above the spin freezing temperature of the CSG phase for 0.4 $> x \geq$ 0.15.

The inconsistency between the exponents $\alpha$, $\beta$ and $\gamma$ can also be understood in light of the interaction between the droplets.
In the present system, there is extremely strong FM fluctuations in the composition range of 0.4 $> x >$ 0.08 as manifested in the large Wilson ratios. \cite{SrCa327HMNF} The interactions between magnetic droplets are then expected to be FM. Therefore the magnetization of the system should grow faster than isolated droplets upon cooling; this leads to a super-Curie temperature dependence, resulting in a power-law behavior with the exponent larger than 1. If this is the case, the evolution of the exponent $\beta$ determined from $M$($T$) should follow the same trend as that of the FM fluctuations, which are exactly observed here. For example, at temperature immediately above $T_f$, the FM fluctuations revealed by the Wilson ratio for $x =$ 0.3 are apparently much weaker than those for $x =$ 0.2 and $ 0.25$, \cite{SrCa327HMNF} and the $\beta$ for $x =$ 0.2 and $ 0.25$ are strikingly larger than that for $x =$ 0.3 (see Fig. \ref{fig:exponents} (b)).
In the high field region, spin fluctuations are suppressed. Thus the exponent $\alpha$ extracted from $M(H)$ does not equal to the exponent $\beta$ derived from $M$($T$). The consistency of the exponent $\gamma$ with $\alpha$ indicates that the electronic specific heat is dominated by the slow dynamics of droplets as expected in the quantum Griffiths model, but less sensitive to the interaction between droplets.

\begin{figure}[t]
\includegraphics[angle=-90,scale=1]{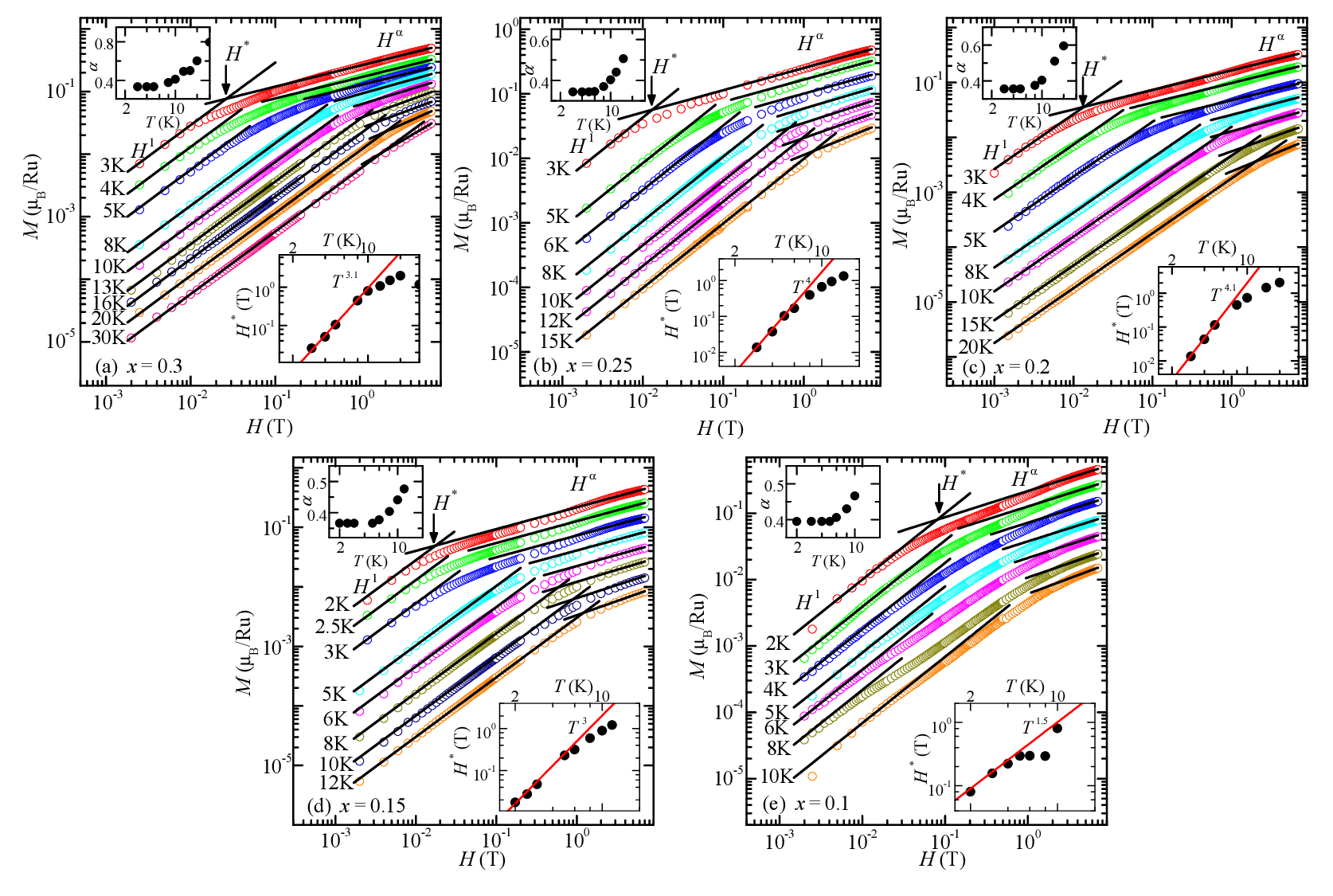}
\caption{(Color online) Isothermal magnetization $M$($H$) measured at typical temperatures for $x =$ 0.3 (a), 0.25 (b), 0.2 (c), 0.15 (d) and 0.1 (e). The magnetic field is applied perpendicular to the $c$-axis. The data are vertically shifted on log-log scale. Solid lines represent the fitting to linear field dependence in low field region or power-law field dependence $H^{\alpha}$ in high field region. $H^{\ast}$ denotes the crossover field. Upper insets show the temperature dependence of the exponent $\alpha$ determined from the power-law fittings. Lower Insets display the temperature dependence of $H^{\ast}$ in log-log scale. The solid lines represent fittings to $T^{-\delta}$.
} \label{fig:MH}
\end{figure}

To further reveal the underlying physics of the anomalous power-law behaviors above $T_{f}$, we have performed the phenomenological scaling analyses of magnetization using the data discussed above.
Given that $M(H)$ shows a linear-to-power-law crossover at a crossover field $H^{\ast}$, analytically we have $\chi(T)H = AH^{\alpha}$ at $H = H^{\ast}$, where $A$ is a constant. Based on this equation, we propose a phenomenological scaling equation $M(H, T) \varpropto H^{\alpha}F(x)$ where $F(x = H/H^{\ast}(T))$ is the scaling function. As discussed above, $\chi = M/H$ is proportional to $T^{-\beta}$. Therefore, it can be shown that $H^{\ast} \varpropto T^{\delta}$ with $\delta = \beta/(1 - \alpha)$. As shown in Fig. \ref{fig:exponents} (b), it is noticed that the values of $\delta$ experimentally determined from $H^{\ast}$($T$) (see insets to Fig. \ref{fig:MH}) approximately equal to those calculated with the equation $\delta = \beta/(1 - \alpha)$, suggesting the self-consistency of the proposed phenomenological scaling law.

The validity of this phenomenological scaling expression is demonstrated in the scaling plots shown in Fig. \ref{fig:scaling} where $M/H^{\alpha}$ is plotted as a function of $H/T^{\delta}$. The magnetization data amazingly collapse into a single curve in a temperature range immediately above $T_{f}$ for the samples with $x$ = 0.3, 0.25, 0.2, and 0.15. The characteristic temperature below which the scaling behavior occurs  is defined as the scaling temperature $T_{scaling}$. At temperatures above $T_{scaling}$, the $M(H, T)$ data deviate from the universal curvature on the $M/H^{\alpha}$ vs. $H/T^{\delta}$ scaling plot. We present an example for the $x$ = 0.3 sample in the inset of Fig. 5a, where the $M(H, T)$ data clearly do not collapse into a universal curve for $T$ $>$ $T_{scaling}$ = 13 K. $T_{scaling}$ systematically decreases with the decease of Ca content.
When the Ca content $x$ decreases to 0.1 near which a CSG-to-IMM phase transition occurs, the scaling law breaks down and the magnetization data clearly does not collapse into a single curve (shown in Fig. \ref{fig:scaling} (e)).
The breakdown of the phenomenological scaling behavior at $x =$ 0.1, as well as the absence of the anomalous power-law behavior of electronic specific heat in this sample, suggest that the underlying physics near this critical composition might be different.
We have added $T_{scaling}$ to the magnetic phase diagram of (Sr$_{1-x}$Ca$_{x}$)$_{3}$Ru$_{2}$O$_{7}$, as shown in Fig. \ref{fig:ps}. Interestingly, $T_{scaling}(x)$ appears to follow the extrapolation of N\'{e}el temperature $T_{N}(x)$, implying that the phenomenological scaling behavior reported here might be a fingerprint of some energy scale in a disordered itinerant system close to a quantum phase transition.

In summary, we have conducted the scaling analyses of magnetization and electronic specific heat for the NFM states in (Sr$_{1-x}$Ca$_{x}$)$_{3}$Ru$_{2}$O$_{7}$. For 0.4 $> x \geq$ 0.15, the isothermal magnetization and the electronic specific heat exhibit anomalous power-law singularity that are controlled by a single exponent in a temperature region immediately above $T_f$. $M$($T$) also follows a power-law behavior. However, the extremely strong FM fluctuations lead its exponent to be different than those derived from the isothermal magnetization and the electronic specific heat. The magnetization data $M(H,T)$ are found to obey a phenomenological scaling law of $M(H,T) \propto H^{\alpha}f(H/T^{\delta})$ also in a temperature region immediately above $T_{f}$. The scaling temperature $T_{scaling}$ systematically decreases with the decease of Ca content. This scaling law, as well as the anomalous power-law behavior, break down near the critical concentration $x =$ 0.1 where a CSG-to-IMM phase transition occurs. These results demonstrate the existence of the slow dynamics of rare regions arising from the effect of disorder on QPT in (Sr$_{1-x}$Ca$_{x}$)$_3$Ru$_2$O$_7$. The finding of the $M(H, T)$ scaling law implies novel physics associated with the slow dynamics of rare regions. Further theoretical work is needed to reveal its nature.

\begin{acknowledgments}
The work at Tulane is supported by the DOD ARO under Grant No. W911NF0910530, the NSF under Grant No. DMR-1205469, and the LA-SiGMA program under Award No. EPS-1003897. Work at HMFL-CAS is supported by National Natural Science Foundation of China under contracts Nos. 11004198 and 11174291. V. D. was supported by NSF through Grant Nos. DMR-0542026 and DMR-1005751. Work at UNO was supported by the DOD ARO under Grant No. W911NF0910530 and the LA-SiGMA program under Award No.EPS-1003897. Z. Q. also gratefully acknowledges supports from the Youth Innovation Promotion Association, Chinese Academy of Sciences.
\end{acknowledgments}

\newpage

\clearpage

\clearpage

\clearpage

\begin{figure}[t]
\includegraphics[angle=-90,scale=1]{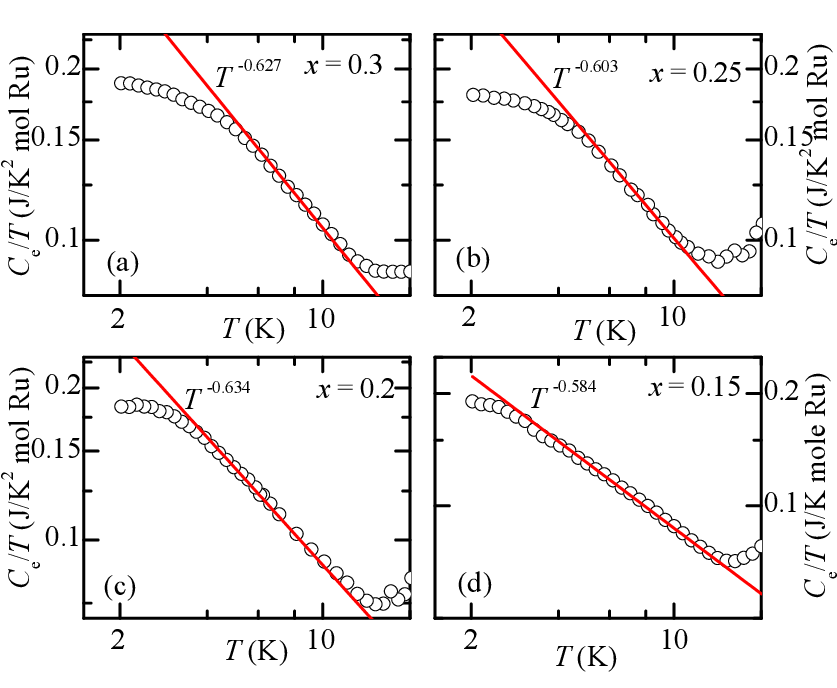}
\caption{(Color online) The electronic specific heat divided by the temperature $C_e/T$ as a function of temperature $T$ for $x =$ 0.3 (a), 0.25 (b), 0.2 (c), and 0.15 (d). The solid lines represent fittings to $T^{-\gamma}$. The data for $x =$ 0.3 are quoted from from our early work. \cite{SrCa327HMNF}}\label{fig:CeT}
\end{figure}

\clearpage

\begin{figure}[t]
\includegraphics[angle=-90,scale=1]{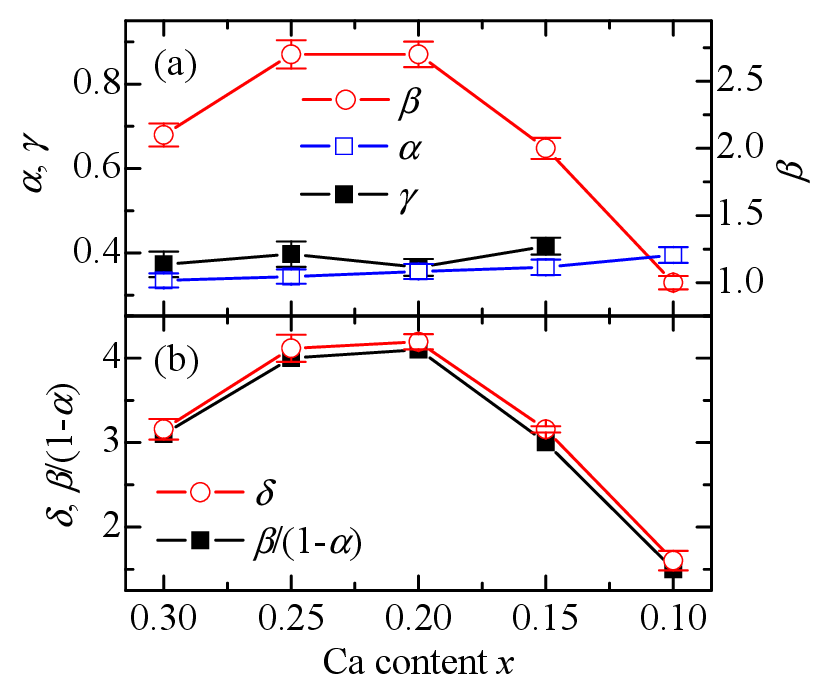}
\caption{(Color online) (a) The exponents $\alpha$, $\beta$, and $\gamma$ as a function of Ca content $x$. (b) The exponents $\delta$ experimentally determined (see lower insets to Fig. \ref{fig:MH}) and calculated using the equation $\delta = \beta/(1-\alpha)$ as a function of Ca content $x$.}\label{fig:exponents}
\end{figure}

\clearpage

\begin{figure}[t]
\includegraphics[angle=-90,scale=1]{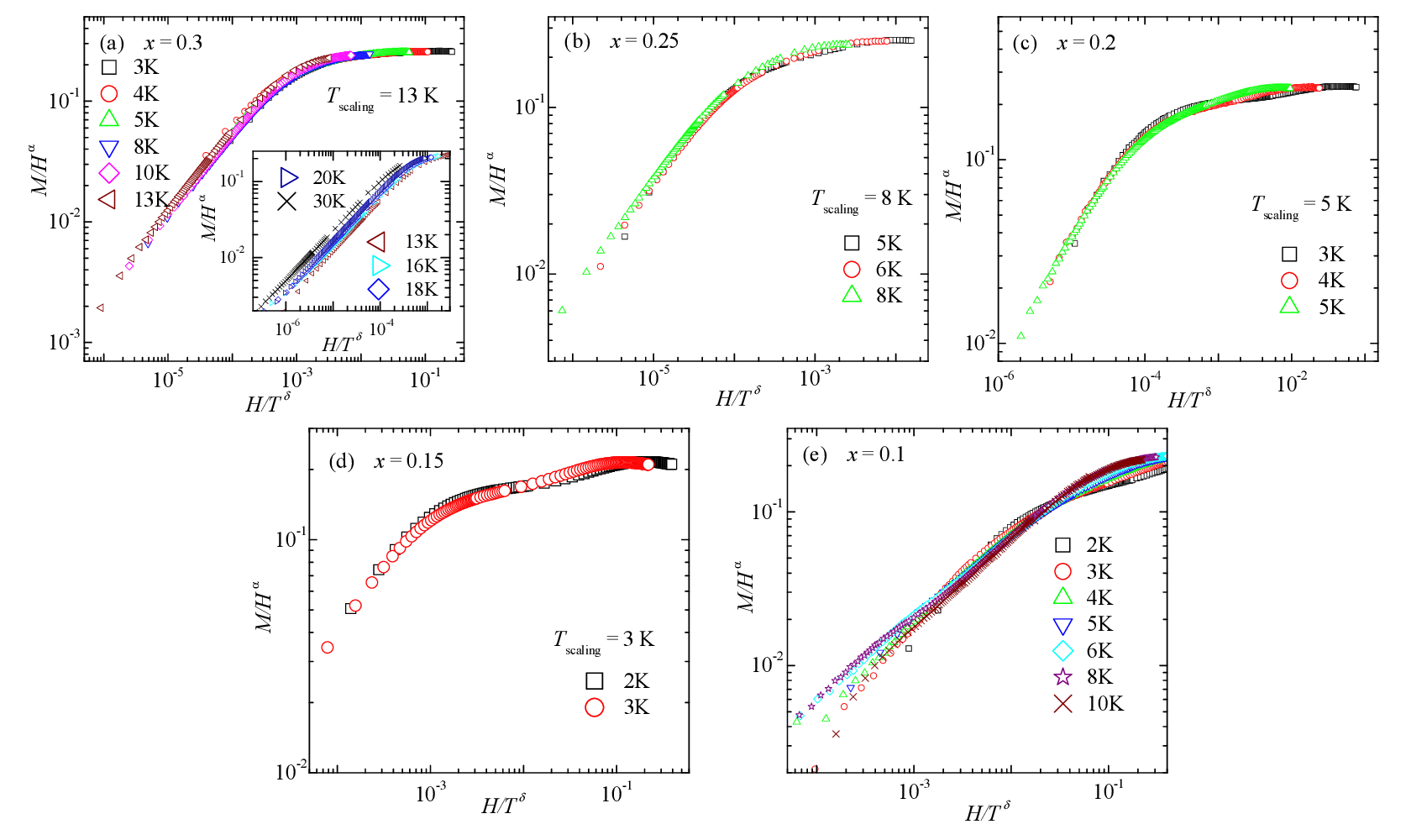}
\caption{(Color online) Scaling plots of $M/H^{\alpha}$ vs. $H/T^{\delta}$ for the samples with $x =$ 0.3 (a), 0.25 (b), 0.2 (c), 0.15 (d) and 0.1 (e). Inset to (a) shows the gradual deviation from the universal curve when the temperature exceeds the scaling temperature $T_{scaling}$.}\label{fig:scaling}
\end{figure}

\clearpage

\begin{figure}[t]
\includegraphics[angle=-90,scale=1]{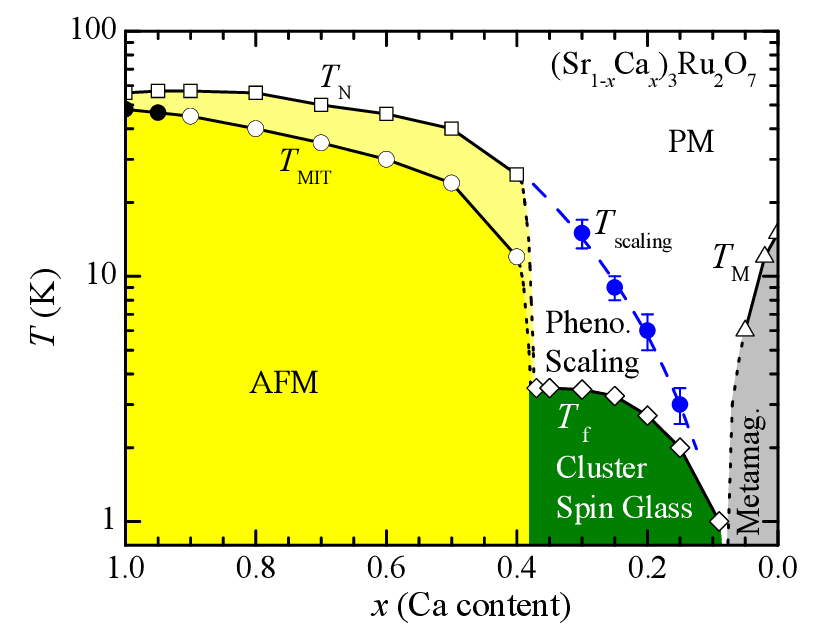}
\caption{(Color online) The updated magnetic phase diagram of (Sr$_{1-x}$Ca$_{x}$)$_{3}$Ru$_{2}$O$_{7}$. \cite{SrCa327HMNF} AFM: antiferromagnetic; $T_{\textrm{N}}$: the N\'{e}el temperature; $T_{\textrm{MIT}}$: the metal-insulator transition temperature; The closed and open circles represent first and second order transition respectively; $T_{\textrm{M}}$: the temperature of the peak in the susceptibility, below which the metamagnetic transition occurs; $T_{\textrm{f}}$: the freezing temperature of the CG phase; $T_{scaling}$: the temperature below which the phenomenological scaling law holds; Metamag.: itinerant metamagnetic; PM: paramagnetic. The dashed lines are for the guide to the eyes.}\label{fig:ps}
\end{figure}

\end{document}